\documentclass[sigconf,natbib=true]{acmart}
\usepackage{enumitem}
\usepackage{subfigure}
\usepackage{multirow}
\usepackage{colortbl}
\newcommand{\wrt}{w.r.t.}
\AtBeginDocument{%
  }

\copyrightyear{2026}
\acmYear{2026}
\setcopyright{cc}
\setcctype{by}
\acmConference[SIGIR '26]{Proceedings of the 49th International ACM SIGIR Conference on Research and Development in Information Retrieval}{July 20--24, 2026}{Melbourne, VIC, Australia}
\acmBooktitle{Proceedings of the 49th International ACM SIGIR Conference on Research and Development in Information Retrieval (SIGIR '26), July 20--24, 2026, Melbourne, VIC, Australia}
\acmDOI{10.1145/3805712.3809861}
\acmISBN{979-8-4007-2599-9/2026/07}

\begin{document}
\title{FEDIN: Frequency-Enhanced Deep Interest Network for Click-Through Rate Prediction}

\author{Zenan Dai}
\authornote{Equal Contribution.}
\orcid{0009-0006-5295-9848}
\affiliation{%
  \institution{Tsinghua University}
  \city{Shenzhen}
  \country{China}
}
\email{dzn24@mails.tsinghua.edu.cn}

\author{Jinpeng Wang}
\authornotemark[1]
\authornote{Corresponding author.}
\orcid{0000-0002-4352-4897}
\affiliation{%
  \institution{Harbin Institute of Technology, Shenzhen}
  \city{Shenzhen}
  \country{China}
}
\email{wangjp26@gmail.com}

\author{Junwei Pan}
\orcid{0009-0003-2697-7012}
\affiliation{%
  \institution{Tencent}
  \city{Shenzhen}
  \country{China}
}
\email{jonaspan@tencent.com}

\author{Dapeng Liu}
\orcid{0009-0003-2973-9167}
\affiliation{%
  \institution{Tencent}
  \city{Shenzhen}
  \country{China}
}
\email{rocliu@tencent.com}

\author{Lei Xiao}
\orcid{0009-0002-3991-8161}
\affiliation{%
  \institution{Tencent}
  \city{Shenzhen}
  \country{China}
}
\email{shawnxiao@tencent.com}

\author{Shu-Tao Xia}
\orcid{0000-0002-8639-982X}
\affiliation{%
  \institution{Tsinghua University}
  \city{Shenzhen}
  \country{China}
}
\email{xiast@sz.tsinghua.edu.cn}

\renewcommand{\shortauthors}{Zenan Dai et al.}

\begin{abstract}
Sequential recommendation models often struggle to capture latent periodic patterns in user interests, primarily due to the noise inherent in time-domain behavioral data. 
While frequency-domain analysis offers a global perspective to address this, existing approaches typically treat user sequences in isolation, overlooking the crucial context of the target item.
In this work, we present a novel empirical observation: user attention scores exhibit distinct spectral entropy distributions when conditioned on positive versus negative target items. 
Specifically, true user interests manifest as highly concentrated spectral patterns with lower entropy in the frequency domain, whereas irrelevant behaviors appear as high-entropy noise.
Leveraging this insight, we propose the Frequency-Enhanced Deep Interest Network (\textbf{FEDIN}). 
FEDIN introduces a frequency-domain branch that utilizes a target-aware spectrum filtering mechanism to isolate these periodic interest signals.
Extensive experiments on three public datasets demonstrate that FEDIN consistently outperforms state-of-the-art sequential recommendation baselines, demonstrating superior robustness against noise.
We have released our code at: \url{https://github.com/otokoneko/FEDIN}.
\end{abstract}


\maketitle

\section{Introduction}
\label{sec: introduction}

Click-Through Rate (CTR) prediction is a core component of modern recommendation systems, serving as a critical bridge between user preferences and candidate items. 
The efficacy of CTR models largely depends on how well they extract user interests from historical behavior sequences~\cite{DIN}. 
While deep sequential models, including RNNs~\cite{DIEN,GRU4Rec,LRURec}, Transformers and SSMs~\cite{BST,SASRec,Bert4Rec,Mamba4Rec,HMamba}, have achieved remarkable success by capturing temporal dependencies, they face inherent challenges in handling real-world user behaviors. 
User history is often a mixture of long-term periodic habits and short-term stochastic noise. 
Purely time-domain approaches, including those with target attention mechanisms~\cite{DIN,DIEN,DSIN,TIN}, tend to be sensitive to such point-wise noise and may struggle to decouple stable periodic patterns from random clicks in long sequences.

\begin{figure}[t]
    \centering
    \includegraphics[width=\linewidth]{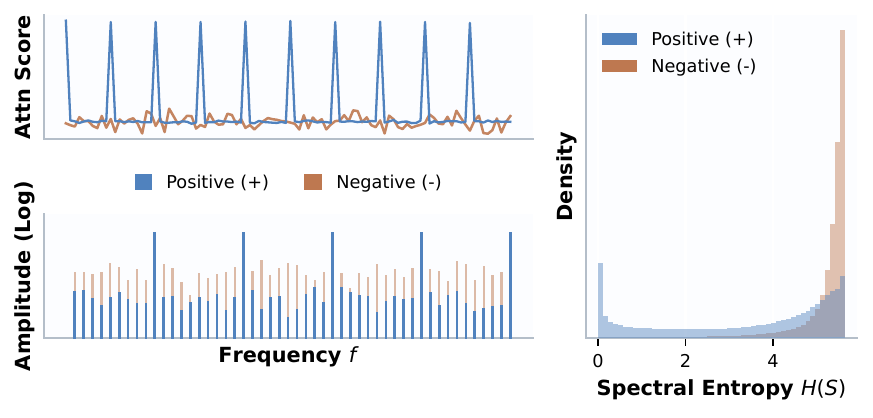}
    \caption{Empirical observation of target-conditioned spectral patterns. (Left) A case study comparing attention scores under different targets: positive items tend to elicit more pronounced periodic resonance in the user sequence, manifesting as concentrated spectral peaks. (Right) Statistical distribution of spectral entropy $H(S) = -\sum p_k \log_2 p_k$ across datasets, showing that positive samples exhibit generally lower entropy signatures compared to negative ones.}
    \vspace{-1em}
    \label{fig:intro_spectral_analysis}
\end{figure}

\begin{figure*}[t!]
	\centering
	\captionsetup{labelfont=bf}
	\subfigure[FEDIN Architecture]{
		\begin{minipage}[b]{0.38\linewidth}
			\label{subfig:FEDIN_overview}
            \includegraphics[width=1\textwidth]{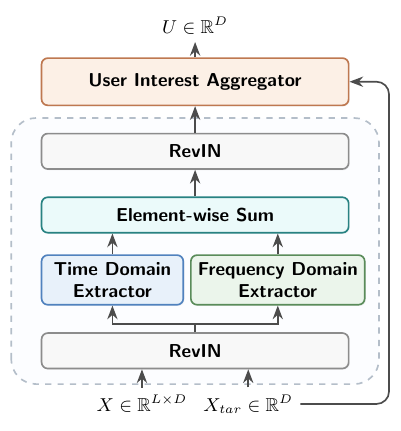}
	\end{minipage}}
	\subfigure[Time Domain Interest Extractor]{
		\begin{minipage}[b]{0.28\linewidth}
			\label{subfig:FEDIN_time}
			\includegraphics[width=1\linewidth]{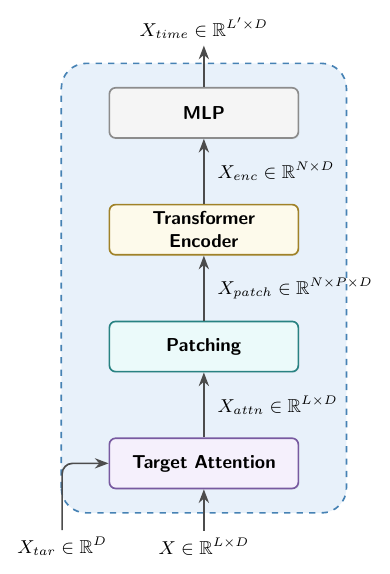}
	\end{minipage}}
	\subfigure[Frequency Domain Interest Extractor]{
		\begin{minipage}[b]{0.28\linewidth}
			\label{subfig:FEDIN_freq}
			\includegraphics[width=1\textwidth]{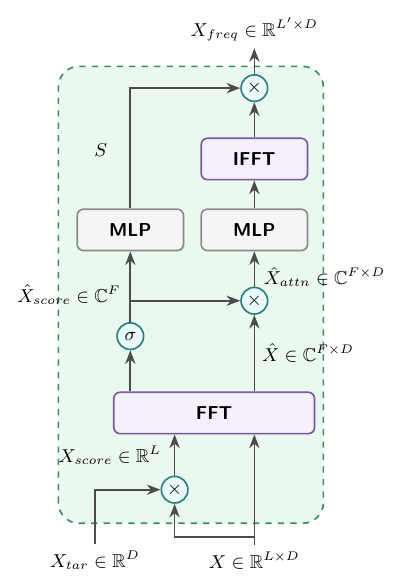}
	\end{minipage}}
	
	\caption{The architecture of FEDIN.}
	\label{fig:network_architecture}
\end{figure*}

To tackle the limitations of time-domain modeling, frequency-domain analysis becomes a promising direction, showing success in long-term time series forecasting~\cite{PDF,FEDformer,FreTS,FiLM}. 
The Fast Fourier Transform offers a global view of signal periodicity, making it naturally suitable for denoising and pattern recognition. 
Inspired by this, recent works have explored frequency-domain analysis to address noise and sparsity. Early attempts like \textbf{FMLP-Rec}~\cite{FilterEnhancedMLP} and \textbf{FEARec}~\cite{FEARec} introduced learnable filters to suppress stochastic noise. More recently, \textbf{DIFF}~\cite{DIFF} has emerged as a state-of-the-art representative, utilizing discrete Fourier transforms to prioritize long-term stable interests through multi-sequence filtering.
However, even advanced models like DIFF apply frequency analysis to the user sequence independently of the candidate item. 
We argue that this target-agnostic filtering is suboptimal because a user's behavior sequence is polymorphic. 
Without the context of a specific target item, the model cannot effectively distinguish between the frequency components representing the true signal and those representing background noise.

In this paper, we revisit frequency-domain recommendation from a target-aware perspective. 
Through a statistical analysis of attention scores derived from a standard target attention backbone, we present a key empirical observation that the Target Attention mechanism exhibits distinct characteristics in the frequency domain. 
When these scores are projected into the frequency domain, true user interests manifest as highly concentrated spectral patterns with lower entropy, whereas irrelevant behaviors exhibit high-entropy distributions resembling white noise.
As shown in Fig.~\ref{fig:intro_spectral_analysis}, when we project the attention scores of user behaviors conditioned on different items into the frequency domain, a clear pattern emerges. 
Specifically, true user interests manifest as highly concentrated spectral patterns with lower entropy, whereas irrelevant behaviors exhibit high-entropy distributions resembling white noise. 
This insight suggests that the target item acts as a natural frequency selector. 
Simply calculating the spectrum of the raw behavior sequence overlooks this critical conditional phenomenon.

Building on this discovery, we propose the Frequency-Enhanced Deep Interest Network (\textbf{FEDIN}). 
Crucially, we recognize that frequency analysis should complement, not replace, time-domain modeling. 
While frequency features capture global periodicities and resist noise, time-domain features excel at capturing delicate local sequential evolution. 
Therefore, FEDIN adopts a dual-branch architecture that effectively fuses these two perspectives. 
The frequency branch utilizes a learnable spectrum filtering mechanism to isolate the structured harmonic patterns identified in our analysis, while the time branch maintains the granularity of sequential interactions. 
This design allows FEDIN to leverage the robustness of the frequency domain without sacrificing the detail of the time domain.

Our contributions are summarized as follows:
\setlist{nolistsep}
\begin{itemize}[leftmargin=1.5em]
\item We empirically identify that user interests exhibit distinct low-entropy signatures in the frequency domain when conditioned on target items, providing a new theoretical grounding for spectral recommendation.
\item We propose FEDIN, a hybrid framework that integrates a target-aware frequency branch with traditional time-domain modeling, effectively combining global noise robustness with local sequential precision.
\item Experiments on three public datasets demonstrate that FEDIN consistently outperforms state-of-the-art sequential baselines, demonstrating superior effectiveness and robustness.
\end{itemize}

\section{Method}
\label{sec:method}

\subsection{Overall Framework}
\label{subsec:overview}

The overall architecture of FEDIN is illustrated in Fig.~\ref{fig:network_architecture}. 
Following the Embedding\&MLP paradigm, FEDIN takes the user behavior sequence $\mathbf{X}\in\mathbb{R}^{L\times D}$ and the target item $\mathbf{X}_{\text{tar}}\in\mathbb{R}^{D}$ as inputs to predict the user interest representation $\mathbf{U}\in\mathbb{R}^{D}$.
Recognizing that user behaviors contain both evolving sequential dependencies and stable periodic patterns, FEDIN employs a dual-branch denoising architecture. 
Notably, this decoupled design allows the two branches to be executed in a highly parallelized manner, ensuring efficiency for real-time deployment.
First, we apply Reversible Instance Normalization (RevIN)~\cite{revin} as a fundamental preprocessing component to mitigate distribution shifts and the non-stationary nature of user behaviors.
The Time-Domain Branch focuses on capturing the local sequential evolution of interests, while the Frequency-Domain Branch focuses on isolating global periodic resonance patterns via spectral filtering.
Finally, the User Interest Aggregator dynamically fuses these views to derive the final representation.

\subsection{Time-Domain Branch}
\label{subsec:time_extractor}

This branch aims to extract interest evolution from the time-domain sequence. 
A naive application of Transformers on raw sequences is often inefficient and does not necessarily guarantee performance gains due to noise~\cite{are_transformers_effective}.
Therefore, we design a coarse-to-fine extraction process.

\textbf{Coarse Filtering via Target Attention.} To mitigate pointwise noise, we first apply a target-aware coarse filter:
\begin{equation}
\label{con:coarse_attn}
    \mathbf{X}_{\text{attn}} = \text{Softmax}(\mathbf{X}_{\text{tar}}\mathbf{X}^\top / \alpha)^\top \odot \mathbf{X},
\end{equation}
where $\odot$ denotes element-wise product. Unlike conventional target attention that performs weighted pooling, we compute a weighted sequence $\mathbf{X}_{\text{attn}} \in \mathbb{R}^{L \times D}$ to preserve temporal resolution for subsequent fine-grained evolution modeling.

\textbf{Fine-grained Evolution via Patching Transformer.}
While TA selects relevant items, it does not model the temporal dependencies among them. 
To capture how these specific interests evolve over time, we employ a Transformer encoder~\cite{vaswani2017attention}.
Crucially, to handle long sequences efficiently and capture local temporal semantics, we adopt a patching strategy~\cite{patchtst}.
Specifically, $\mathbf{X}_{\text{attn}}$ is segmented into non-overlapping patches $\mathbf{X}_{\text{patch}}\in\mathbb{R}^{N\times P\times D}$. 
We pad the sequence with zeros if $L$ is not divisible by $P$, then flatten and linearly project each patch to dimension $D$.
A Transformer encoder then models the dependencies between these patches, yielding the time-domain interest representation $\mathbf{X}_{\text{time}}$.
This design ensures that we not only identify what the user likes but also how that interest changes over time.

\subsection{Frequency-Domain Branch}
\label{subsec:freq_extractor}

As discussed in Section~\ref{sec: introduction}, true user interests manifest as concentrated harmonic patterns in the frequency domain.
This branch aims to recover these signals from noisy behaviors using a target-aware spectral filtering mechanism.

First, we calculate the raw target attention scores $\mathbf{X}_{\text{score}} \in \mathbb{R}^{L}$ to capture the relevance of each historical behavior to the target item:
\begin{equation}
\label{con:score}
    \mathbf{X}_{\text{score}} = \mathbf{X}\mathbf{X}_{\text{tar}}^\top / \alpha,
\end{equation}
where $\alpha$ is the scaling factor. Let $\hat{\mathbf{X}}_{\text{score}} = \mathcal{F}(\mathbf{X}_{\text{score}}) \in \mathbb{C}^{L}$ represent the target-conditioned spectrum. Unlike prior works that transform the raw sequence $\mathbf{X}$, we transform these attention scores to ensure the spectrum is conditioned on the target context. The target-aware spectrum is computed as:
\begin{equation}
\label{con:FTA_score}
    \hat{\mathbf{X}}_{\text{attn}} = \text{Softmax}(\text{Amp}(\hat{\mathbf{X}}_{\text{score}})) \otimes \mathcal{F}(\mathbf{X}),
\end{equation}
where $\mathcal{F}(\cdot)$ is the FFT operation and $\otimes$ denotes the broadcasting element-wise product along the dimension $D$. This operation highlights frequency components where the user shows strong resonance with the target item.

\textbf{Learnable Spectral Filter.}
To isolate the signal from noise, we need a filter to suppress high-entropy frequencies. 
Instead of using fixed filters, we employ a Complex-Valued MLP as a learnable spectral filter, adhering to the standard complex formulation defined in~\cite{trabelsi2017deep} to strictly preserve the intrinsic phase-amplitude coupling.
The MLP acts on the complex spectrum, adaptively amplifying dominant harmonic frequencies while suppressing noise:
\begin{equation}
    \hat{\mathbf{X}}_{\text{freq}}=\mathcal{F}^{-1}(\text{MLP}(\hat{\mathbf{X}}_{\text{attn}})).
\end{equation}
Here, the MLP learns the non-linear mapping from the noisy input spectrum to the clean interest spectrum.

\textbf{Adaptive Resonance Scaling.}
Not all user-item pairs exhibit strong periodicity.
Based on our empirical observation, strong interests show lower spectral entropy.
To leverage this, we introduce an adaptive gating mechanism that scales the frequency branch output based on the spectral clarity:
\begin{equation}
\label{con:FT_mixer}
    \mathbf{X}_{\text{freq}}=\text{Sigmoid}(\text{MLP}(\text{Amp}(\hat{\mathbf{X}}_{\text{score}})))\cdot\hat{\mathbf{X}}_{\text{freq}}.
\end{equation}
This ensures the model relies more on the frequency view when clear resonance patterns are detected, and falls back to the time domain otherwise.

\textbf{Complexity Analysis.} The computational cost of the Frequency-Domain Branch is dominated by FFT operations with a complexity of $O(L \log L)$, where $L$ is the sequence length. This provides a significant efficiency advantage over the $O(L^2)$ complexity of standard self-attention mechanisms, particularly as $L$ increases.

\subsection{User Interest Aggregator}
\label{subsec:ui_agg}

The two branches provide complementary views, where $\mathbf{X}_{\text{time}}$ captures local evolution and $\mathbf{X}_{\text{freq}}$ captures global periodicity.
We sum them to obtain the mixed representation $\mathbf{X}_{\text{mix}}$.
To derive the final user interest $\mathbf{U}$, we employ a Top-k Target Attention mechanism:
\begin{equation}
    \mathbf{U}=\text{Softmax}(\text{Top-k}(\mathbf{X}_{\text{tar}}\mathbf{X}_{\text{mix}}^\top/\alpha,k)) \cdot \mathbf{X}_{\text{mix}}.
\end{equation}
Unlike conventional target attention, the top-k target attention retains only the $k$ items with the highest similarity.
To ensure differentiability, we implement this by masking the non-top-$k$ attention scores to $-\infty$ before the Softmax operation, allowing gradients to flow through the selected indices.
This design draws inspiration from multi-task learning strategies that isolate parameters to prevent the seesaw phenomenon~\cite{mmoe,PLE,STEM}, ensuring that conflicting or weak interest signals do not dilute the final prediction.

\section{Experiments}
\label{sec:experiments}

\subsection{Experimental Setup}
\label{subsec:setup}
\noindent\textbf{Datasets \& Protocols.} 
We evaluate FEDIN on three public datasets: \textbf{Tmall}\footnote{\url{https://tianchi.aliyun.com/dataset/dataDetail?dataId=42}}, \textbf{Taobao}\footnote{\url{https://tianchi.aliyun.com/dataset/dataDetail?dataId=649}}, and \textbf{Alipay}\footnote{\url{https://tianchi.aliyun.com/dataset/dataDetail?dataId=53}}. 
The datasets are partitioned into training, validation, and testing sets based on the global timeline.
For evaluation, we employ \textbf{AUC} and \textbf{Group AUC (GAUC)} as the primary metrics~\cite{DIN}.

\noindent\textbf{Baselines.}
To comprehensively evaluate the effectiveness of our proposed FEDIN model, we conducted a thorough comparison with six state-of-the-art CTR prediction baseline models, including \textbf{DIN}~\cite{DIN}, \textbf{DIEN}~\cite{DIEN} ,\textbf{SASRec}~\cite{SASRec}, \textbf{BERT4Rec}~\cite{Bert4Rec}, \textbf{GRU4Rec}~\cite{GRU4Rec}, \textbf{BST}~\cite{BST} and \textbf{DIFF}~\cite{DIFF}. In addition, we utilize a Sum Pooling model as the base model, which transforms the list of user behavior embedding vectors into a fixed-length feature vector through sum pooling. Notably, while DIFF was originally designed for the next-item prediction task, we adapt it for the target-aware CTR task by replacing its ranking-based output layer with an MLP prediction head that incorporates the candidate item embedding.

\noindent\textbf{Implementation Details.}
To ensure fair comparison, all models are implemented based on FuxiCTR~\cite{fuxictr}. 
We use the Adam optimizer with a learning rate of $5e^{-4}$ and a batch size of 2048. 
The embedding dimension is set to 32, and the maximum sequence length is fixed at 100.

\subsection{Overall Performance}
\label{subsec:ovearll_perf}

\begin{table}[t]
	\caption{The overall performance of various methods on three publicly available recommendation datasets. Bold indicates the best performance, while underlined denotes the best performance of the baseline model. An asterisk (*) implies the improvements over the best baseline are statistically significant ($p$--value < 0.05).}
	\label{tab:exp}
    \resizebox{\columnwidth}{!}{
    	\begin{tabular}{l|cc|cc|cc}
    		\toprule[1.3pt]
    		\multirow{2}{*}{Model} & \multicolumn{2}{c}{Tmall} & \multicolumn{2}{|c}{Alipay} & \multicolumn{2}{|c}{Taobao}\\
    		
    		\cmidrule(lr){2-3} \cmidrule(lr){4-5} \cmidrule(lr){6-7}
    		& GAUC & AUC  & GAUC & AUC & GAUC & AUC  \\
    		\midrule
    		Sum Pooling & 0.8978 & 0.8873 & 0.8557 & 0.8535 & 0.9345 & 0.9337\\
    		DIN         & \underline{0.9547} & \underline{0.9518} & 0.8897 & 0.8832 & \underline{0.9689} & \underline{0.9664}\\
    		DIEN        & 0.9237 & 0.9157 & 0.8980 & 0.8953 & 0.9443 & 0.9442\\
    		SASRec      & 0.9183 & 0.9246 & 0.9238 & \underline{0.9312} & 0.9583 & 0.9584\\
    		BERT4Rec    & 0.9103 & 0.9157 & 0.9179 & 0.9189 & 0.9523 & 0.9535\\
    		GRU4Rec     & 0.9210 & 0.9242 & \underline{0.9268} & 0.9289 & 0.9618 & 0.9638\\
    		BST         & 0.9233 & 0.9269 & 0.9264 & 0.9285 & 0.9562 & 0.9576\\
    		DIFF         & 0.8513 & 0.8618 & 0.8962 & 0.8995 & 0.9463 & 0.9475\\
    		\midrule
    		
    		\rowcolor{green!10} FEDIN (Ours)          & \textbf{0.9658\text{*}} & \textbf{0.9666\text{*}} & \textbf{0.9335\text{*}} & \textbf{0.9320\text{*}} & \textbf{0.9740\text{*}} & \textbf{0.9729\text{*}}\\
    	
    		\bottomrule[1.3pt]
    	\end{tabular}
    }
        \vspace{-0.8em}
\end{table}

The quantitative results are presented in Table~\ref{tab:exp}. 
FEDIN consistently outperforms the best-performing baselines across all three datasets, validating the efficacy of our frequency-enhanced approach.
We observe that baseline models exhibit varying degrees of effectiveness depending on the dataset characteristics. 
For instance, attention-based models like DIN perform exceptionally well on Tmall and Taobao but struggle on Alipay, likely due to differences in data sparsity and behavior complexity.
In contrast, FEDIN demonstrates superior robustness, achieving stable state-of-the-art performance in all scenarios.
This stability can be attributed to the dual-branch architecture, where the frequency domain branch effectively compensates for the instability of time-domain features in sparse or noisy environments.

\subsection{Ablation Study}
\label{subsec:ablation_study}

\begin{table}[t]
	\caption{Ablation analysis on three datasets. The best performance is denoted in bold.}
	\label{tab:exp_ablation}
    \resizebox{\columnwidth}{!}{
    	\begin{tabular}{l|cc|cc|cc}
    		\toprule[1.3pt]
    		\multirow{2}{*}{Model} & \multicolumn{2}{c}{Tmall} & \multicolumn{2}{|c}{Alipay} & \multicolumn{2}{|c}{Taobao}\\
    		
    		\cmidrule(lr){2-3} \cmidrule(lr){4-5} \cmidrule(lr){6-7}
    		& GAUC & AUC  & GAUC & AUC & GAUC & AUC  \\
    		\midrule
    		\rowcolor{green!10} FEDIN          & \textbf{0.9658} & \textbf{0.9666} & \textbf{0.9335} & \textbf{0.9320} & \textbf{0.9740} & \textbf{0.9729}\\
            
    		w/o Time-Domain Branch & 0.9482 & 0.9498 & 0.9253 & 0.9246 & 0.9674 & 0.9669\\
    		w/o Freq-Domain Branch & 0.9605 & 0.9618 & 0.9270 & 0.9263 & 0.9727 & 0.9715\\
    		w/o Freq-Domain TA & 0.9614 & 0.9626 & 0.9327 & 0.9293 & 0.9732 & 0.9724\\
    		w/o Freq-Domain Scaling & 0.9625 & 0.9636 & 0.9303 & 0.9296 & 0.9662 & 0.9655\\
    	
    		\bottomrule[1.3pt]
    	\end{tabular}
    }
\end{table}

To analyze the necessity of each component in the proposed model, we conducted ablation experiments. Table~\ref{tab:exp_ablation} presents the performance of FEDIN and its variants across three datasets. Overall, the removal of any component leads to a degradation in performance.

\vspace{0.3em}\noindent\textbf{Impact of Dual-Branch Architecture.}
We first evaluate the necessity of integrating both time and frequency domains. 
Removing the Time-Domain Branch (w/o Time-Domain Branch) leads to a significant performance drop, confirming that local sequential evolution is crucial for immediate interest prediction. 
Similarly, removing the Frequency-Domain Branch (w/o Freq-Domain Branch) also results in degradation. 
This proves that frequency-domain features capture complementary information—specifically global periodic patterns—that time-domain models alone fail to overlook.

\vspace{0.3em}\noindent\textbf{Impact of Target-Aware Spectral Filtering.}
A key innovation of FEDIN is the incorporation of target items into the frequency analysis. 
In the w/o Freq-Domain TA variant, we replaced the target-aware spectrum generation with a standard learnable frequency filter. 
Specifically, we replaced the dynamic calculation of the attention score spectrum in Eqn.~\eqref{con:FTA_score} with a static, learnable weight matrix.
This matrix directly filters the user behavior spectrum, decoupling the frequency analysis from the target item.
The decline in metrics validates our core hypothesis: simply analyzing the user's behavior spectrum is insufficient. 
The target item acts as a necessary condition to activate specific resonance patterns, allowing the model to distinguish relevant signals from background noise.

\vspace{0.3em}\noindent\textbf{Impact of Adaptive Scaling.}
The w/o Freq-Domain Scaling variant removes the adaptive gating mechanism. 
In this configuration, we omitted the scaling factor described in Eqn.~\eqref{con:FT_mixer} and directly used the unscaled output of the spectral filter as the branch output.
The observed performance gap suggests that not all user-item interactions exhibit strong periodicity. 
The adaptive scaling allows the model to dynamically adjust the reliance on frequency features, prioritizing them only when clear spectral structures are detected, thereby enhancing model flexibility.

\subsection{Hyperparameter Analysis}
\label{subsec:hyperparameter}

\begin{figure}[t]
	\centering
	\captionsetup{labelfont=bf}
	\subfigure[Performances \wrt Patch Size.]{
		\begin{minipage}{0.45\linewidth}
			\label{subfig:param_patch_size}
   \includegraphics[width=\linewidth]{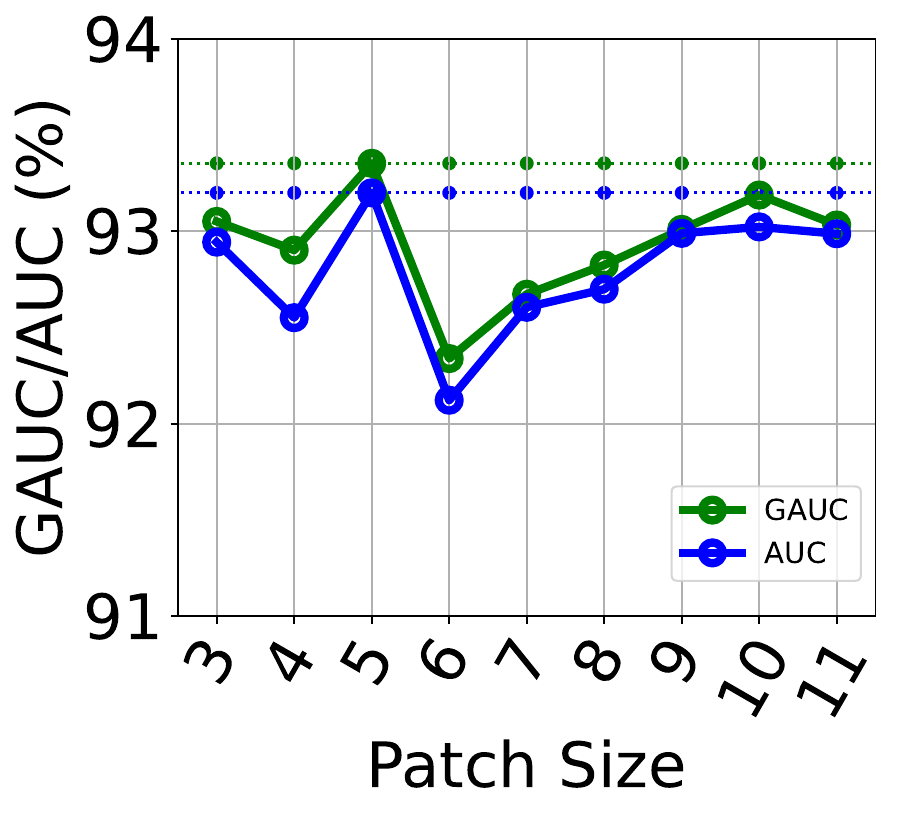}
	\end{minipage}} \quad
	\subfigure[Performances \wrt different K.]{
		\begin{minipage}{0.45\linewidth}
			\label{subfig:param_k}
   \includegraphics[width=\linewidth]{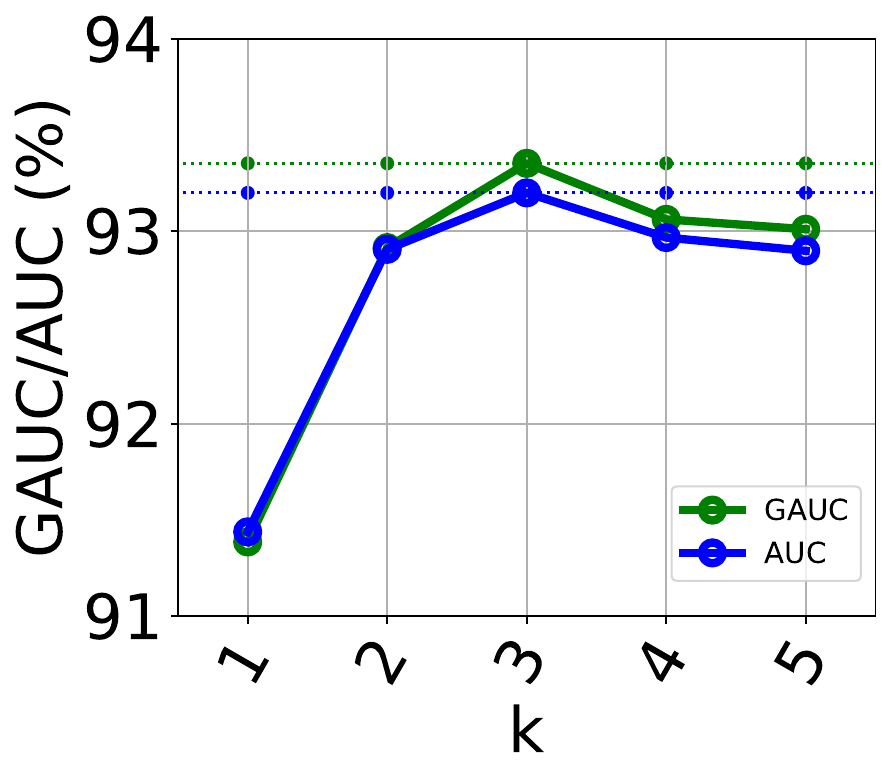}
	\end{minipage}}
	\caption{Hyperparameter analysis on the Taobao dataset.}
        \vspace{-2em}
	\label{fig:params}
\end{figure}

There are two hyperparameters of FEDIN that significantly impact performance: 1) the size of each patch in the time-domain branch, denoted as Patch Size; and 2) the number of candidate interest vectors to retain in the User Interest Aggregator, denoted as K. Fig.~\ref{fig:params} illustrates the results of our experiments on public datasets to determine the optimal settings. For Patch Size, extreme values either fragment local context or cause information loss. 
For $K$, a balance is required between interest diversity and noise reduction; a small $K$ limits coverage, while a large $K$ introduces irrelevant signals and exacerbates the seesaw phenomenon.

\subsection{Noise Resistance Ability Analysis}
\label{subsec:noise_resistance}

\begin{figure}[t]
	\centering
	\captionsetup{labelfont=bf}
	\subfigure[Relative Performance \wrt Drop Noise.]{
		\begin{minipage}{0.45\linewidth}
			\label{subfig:noise_drop}
   \includegraphics[width=\linewidth]{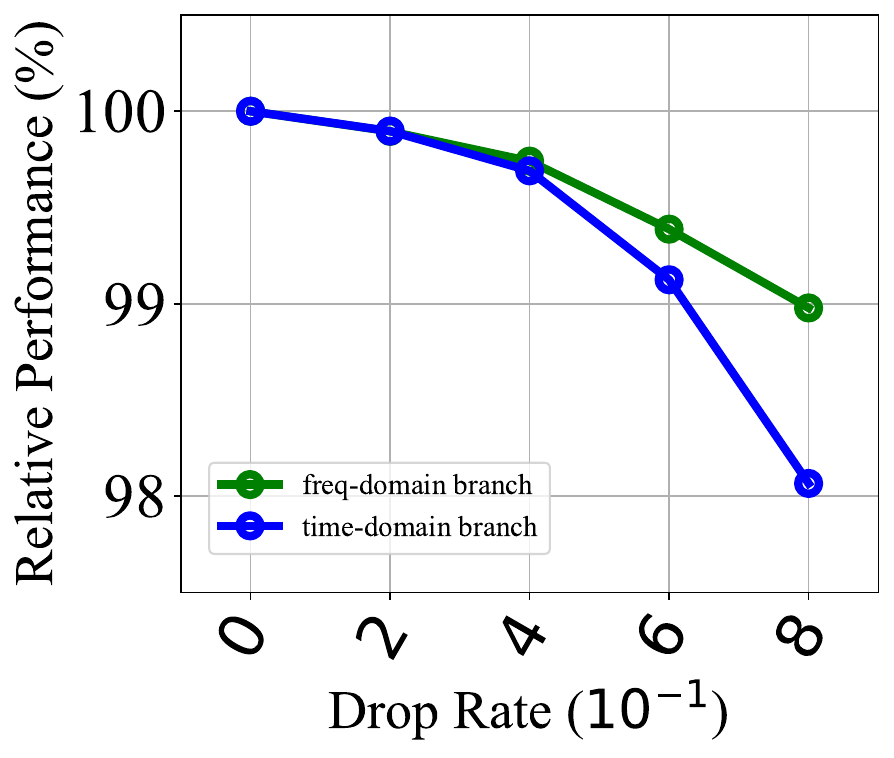}
	\end{minipage}} \quad
	\subfigure[Relative Performance \wrt Replace Noise.]{
		\begin{minipage}{0.45\linewidth}
			\label{subfig:noise_replace}
   \includegraphics[width=\linewidth]{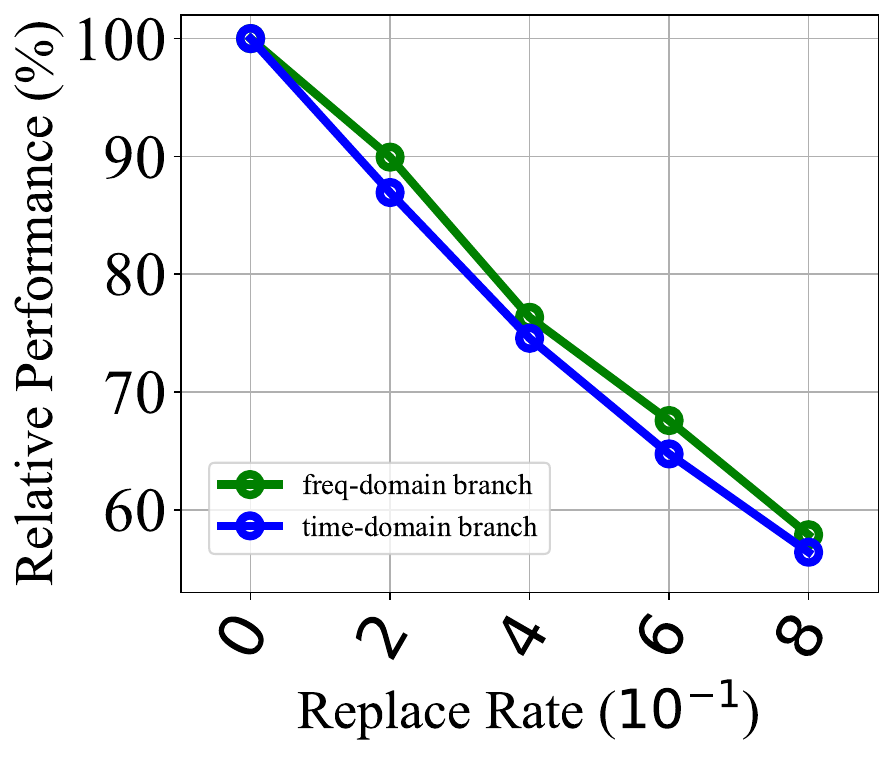}
	\end{minipage}}
	\caption{Noise resistance analysis on the Taobao dataset.}
	\label{fig:noise}
\end{figure}

To further validate the robustness of our frequency-enhanced design, we conducted a synthetic noise resistance experiment. Since real-world interaction logs naturally suffer from sparsity and accidental clicks, we simulated two corresponding noise scenarios at varying corruption ratios ($\rho \in \{0.0, 0.2, 0.4, 0.6, 0.8\}$): (1) \textbf{Drop Noise}, which randomly deletes historical behaviors to mimic missing data, and (2) \textbf{Replace Noise}, which substitutes true behaviors with uniformly sampled random items to mimic exploratory or accidental clicks. We then compared the relative performance decay of the standalone Time-Domain Branch versus the Frequency-Domain Branch.

As shown in Fig.~\ref{fig:noise}, the Frequency-Domain Branch demonstrates significantly higher stability as the noise rate $\rho$ increases. 
In the time domain, deleting or replacing even a few items severely disrupts the local sequential context, often causing error propagation in standard RNNs or Transformers. 
Conversely, in the frequency domain, random noise typically manifests as dispersed, high-entropy fluctuations, whereas core user interests concentrate energy into distinct resonant peaks. 
By applying spectral filtering, FEDIN effectively attenuates these high-entropy noise components and reconstructs the underlying interest pattern from the corrupted raw sequence. 
This analysis confirms that integrating frequency-domain modeling significantly enhances the robustness of recommendation systems against real-world data imperfections.
\section{Conclusions}
\label{sec:conclusion}

This study explored methods for further applying frequency domain analysis techniques in CTR prediction. To facilitate this research, we propose FEDIN, which extracts frequency domain features by incorporating a target attention mechanism and integrates both time domain and frequency domain analysis methods. This allows FEDIN to maintain stable performance across datasets with varying distributions. Comprehensive experiments conducted on three public datasets demonstrate FEDIN's effectiveness.

\begin{acks}
We want to thank the anonymous reviewers and the meta-reviewer for their valuable comments and suggestions. 
This work is supported in part by the National Natural Science Foundation of China under Grants 62571298 and 624B2088.
\end{acks}

\bibliographystyle{ACM-Reference-Format}
\bibliography{main}

@inproceedings{vaswani2017attention,
  author       = {Ashish Vaswani and Noam Shazeer and Niki Parmar and Jakob Uszkoreit and Llion Jones and Aidan N. Gomez and Lukasz Kaiser and Illia Polosukhin},
  title        = {Attention is All you Need},
  booktitle    = {Advances in Neural Information Processing Systems 30},
  pages        = {5998--6008},
  year         = {2017}
}

@inproceedings{DIN,
  author       = {Guorui Zhou and others},
  title        = {Deep Interest Network for Click-Through Rate Prediction},
  booktitle    = {Proc. KDD},
  pages        = {1059--1068},
  year         = {2018}
}

@inproceedings{DIEN,
  author       = {Guorui Zhou and Na Mou and Ying Fan and Qi Pi and Weijie Bian and Chang Zhou and Xiaoqiang Zhu and Kun Gai},
  title        = {Deep Interest Evolution Network for Click-Through Rate Prediction},
  booktitle    = {The Thirty-Third {AAAI} Conference on Artificial Intelligence},
  pages        = {5941--5948},
  year         = {2019}
}

@inproceedings{GRU4Rec,
  author       = {Bal{\'{a}}zs Hidasi and Alexandros Karatzoglou and Linas Baltrunas and Domonkos Tikk},
  title        = {Session-based Recommendations with Recurrent Neural Networks},
  booktitle    = {4th International Conference on Learning Representations, {ICLR} 2016},
  year         = {2016}
}

@article{BST,
  author       = {Qiwei Chen and Huan Zhao and Wei Li and Pipei Huang and Wenwu Ou},
  title        = {Behavior Sequence Transformer for E-commerce Recommendation in Alibaba},
  journal      = {CoRR},
  volume       = {abs/1905.06874},
  year         = {2019}
}

@inproceedings{SASRec,
  author       = {Wang{-}Cheng Kang and Julian J. McAuley},
  title        = {Self-Attentive Sequential Recommendation},
  booktitle    = {{IEEE} International Conference on Data Mining, {ICDM} 2018},
  pages        = {197--206},
  year         = {2018}
}

@inproceedings{Bert4Rec,
  author       = {Fei Sun and Jun Liu and Jian Wu and Changhua Pei and Xiao Lin and Wenwu Ou and Peng Jiang},
  title        = {{BERT4Rec:} Sequential Recommendation with Bidirectional Encoder Representations from Transformer},
  booktitle    = {Proceedings of the 28th {ACM} International Conference on Information and Knowledge Management, {CIKM} 2019},
  pages        = {1441--1450},
  year         = {2019}
}

@inproceedings{PDF,
  author       = {Tao Dai and Beiliang Wu and Peiyuan Liu and Naiqi Li and Jigang Bao and Yong Jiang and Shu{-}Tao Xia},
  title        = {Periodicity Decoupling Framework for Long-term Series Forecasting},
  booktitle    = {The Twelfth International Conference on Learning Representations, {ICLR} 2024},
  year         = {2024}
}

@inproceedings{FEDformer,
  author       = {Tian Zhou and Ziqing Ma and Qingsong Wen and Xue Wang and Liang Sun and Rong Jin},
  title        = {{FEDformer:} Frequency Enhanced Decomposed Transformer for Long-term Series Forecasting},
  booktitle    = {International Conference on Machine Learning, {ICML} 2022},
  series       = {Proceedings of Machine Learning Research},
  volume       = {162},
  pages        = {27268--27286},
  year         = {2022}
}

@inproceedings{FilterEnhancedMLP,
  author       = {Kun Zhou and Hui Yu and Wayne Xin Zhao and Ji{-}Rong Wen},
  title        = {Filter-enhanced {MLP} is All You Need for Sequential Recommendation},
  booktitle    = {The {ACM} Web Conference 2022},
  pages        = {2388--2399},
  year         = {2022}
}

@inproceedings{revin,
  author       = {Taesung Kim and Jinhee Kim and Yunwon Tae and Cheonbok Park and Jang{-}Ho Choi and Jaegul Choo},
  title        = {Reversible Instance Normalization for Accurate Time-Series Forecasting against Distribution Shift},
  booktitle    = {The Tenth International Conference on Learning Representations, {ICLR} 2022},
  year         = {2022}
}

@inproceedings{patchtst,
  author       = {Yuqi Nie and Nam H. Nguyen and Phanwadee Sinthong and Jayant Kalagnanam},
  title        = {A Time Series is Worth 64 Words: Long-term Forecasting with Transformers},
  booktitle    = {The Eleventh International Conference on Learning Representations, {ICLR} 2023},
  year         = {2023}
}

@inproceedings{are_transformers_effective,
  author       = {Ailing Zeng and Muxi Chen and Lei Zhang and Qiang Xu},
  title        = {Are Transformers Effective for Time Series Forecasting?},
  booktitle    = {Thirty-Seventh {AAAI} Conference on Artificial Intelligence, {AAAI} 2023},
  pages        = {11121--11128},
  year         = {2023}
}

@inproceedings{mmoe,
  author       = {Jiaqi Ma and Zhe Zhao and Xinyang Yi and Jilin Chen and Lichan Hong and Ed H. Chi},
  title        = {Modeling Task Relationships in Multi-task Learning with Multi-gate Mixture-of-Experts},
  booktitle    = {Proceedings of the 24th {ACM} {SIGKDD} International Conference on Knowledge Discovery {\&} Data Mining},
  pages        = {1930--1939},
  year         = {2018}
}

@inproceedings{fuxictr,
  author       = {Jieming Zhu and Jinyang Liu and Shuai Yang and Qi Zhang and Xiuqiang He},
  title        = {Open Benchmarking for Click-Through Rate Prediction},
  booktitle    = {The 30th {ACM} International Conference on Information and Knowledge Management, {CIKM} '21},
  pages        = {2759--2769},
  year         = {2021}
}

@inproceedings{FEARec,
  author       = {Xinyu Du and Huanhuan Yuan and Pengpeng Zhao and Jianfeng Qu and Fuzhen Zhuang and Guanfeng Liu and Yanchi Liu and Victor S. Sheng},
  title        = {Frequency Enhanced Hybrid Attention Network for Sequential Recommendation},
  booktitle    = {Proceedings of the 46th International {ACM} {SIGIR} Conference on Research and Development in Information Retrieval},
  pages        = {78--88},
  year         = {2023}
}

@inproceedings{TIN,
  author       = {Haolin Zhou and Junwei Pan and Xinyi Zhou and Xihua Chen and Jie Jiang and Xiaofeng Gao and Guihai Chen},
  title        = {Temporal Interest Network for User Response Prediction},
  booktitle    = {Companion Proceedings of the {ACM} on Web Conference 2024, {WWW} 2024},
  pages        = {413--422},
  year         = {2024}
}

@article{Mamba4Rec,
  author       = {Chengkai Liu and Jianghao Lin and Jianling Wang and Hanzhou Liu and James Caverlee},
  title        = {{Mamba4Rec:} Towards Efficient Sequential Recommendation with Selective State Space Models},
  journal      = {CoRR},
  volume       = {abs/2403.03900},
  year         = {2024}
}

@inproceedings{STEM,
  author       = {Liangcai Su and Junwei Pan and Ximei Wang and Xi Xiao and Shijie Quan and Xihua Chen and Jie Jiang},
  title        = {{STEM:} Unleashing the Power of Embeddings for Multi-Task Recommendation},
  booktitle    = {Thirty-Eighth {AAAI} Conference on Artificial Intelligence, {AAAI} 2024},
  pages        = {9002--9010},
  year         = {2024}
}

@inproceedings{PLE,
  author       = {Hongyan Tang and Junning Liu and Ming Zhao and Xudong Gong},
  title        = {Progressive Layered Extraction {(PLE):} {A} Novel Multi-Task Learning {(MTL)} Model for Personalized Recommendations},
  booktitle    = {Fourteenth {ACM} Conference on Recommender Systems, RecSys 2020},
  pages        = {269--278},
  year         = {2020}
}

@inproceedings{LRURec,
  author       = {Zhenrui Yue and Yueqi Wang and Zhankui He and Huimin Zeng and Julian J. McAuley and Dong Wang},
  title        = {Linear Recurrent Units for Sequential Recommendation},
  booktitle    = {Proceedings of the 17th {ACM} International Conference on Web Search and Data Mining, {WSDM} 2024},
  pages        = {930--938},
  year         = {2024}
}

@inproceedings{DSIN,
  author       = {Yufei Feng and Fuyu Lv and Weichen Shen and Menghan Wang and Fei Sun and Yu Zhu and Keping Yang},
  title        = {Deep Session Interest Network for Click-Through Rate Prediction},
  booktitle    = {Proceedings of the Twenty-Eighth International Joint Conference on Artificial Intelligence, {IJCAI} 2019},
  pages        = {2301--2307},
  year         = {2019}
}

@inproceedings{FreTS,
  author       = {Kun Yi and Qi Zhang and Wei Fan and Shoujin Wang and Pengyang Wang and Hui He and Ning An and Defu Lian and Longbing Cao and Zhendong Niu},
  title        = {Frequency-domain MLPs are More Effective Learners in Time Series Forecasting},
  booktitle    = {Advances in Neural Information Processing Systems 36, NeurIPS 2023},
  year         = {2023}
}

@inproceedings{FiLM,
  author       = {Tian Zhou and Ziqing Ma and Xue Wang and Qingsong Wen and Liang Sun and Tao Yao and Wotao Yin and Rong Jin},
  title        = {{FiLM:} Frequency improved Legendre Memory Model for Long-term Time Series Forecasting},
  booktitle    = {Advances in Neural Information Processing Systems 35, NeurIPS 2022},
  year         = {2022}
}

@article{trabelsi2017deep,
  title={Deep complex networks},
  author={Trabelsi, Chiheb and Bilaniuk, Olexa and Zhang, Ying and Serdyuk, Dmitriy and Subramanian, Sandeep and Santos, Joao Felipe and Mehri, Soroush and Rostamzadeh, Negar and Bengio, Yoshua and Pal, Christopher J},
  journal={arXiv preprint arXiv:1705.09792},
  year={2017}
}

@inproceedings{DIFF,
  title={DIFF: Dual Side-Information Filtering and Fusion for Sequential Recommendation},
  author={Kim, Hye-young and Choi, Minjin and Lee, Sunkyung and Baek, Ilwoong and Lee, Jongwuk},
  booktitle={Proceedings of the 48th International ACM SIGIR Conference on Research and Development in Information Retrieval},
  pages={1624--1633},
  year={2025}
}

@article{HMamba,
  title={HMamba: Hyperbolic Mamba for Sequential Recommendation},
  author={Zhang, Qianru and Wen, Honggang and Yuan, Wei and Chen, Crystal and Yang, Menglin and Yiu, Siu-Ming and Yin, Hongzhi},
  journal={arXiv preprint arXiv:2505.09205},
  year={2025}
}

\end{document}